\documentclass[letterpaper, 10 pt, conference]{ieeeconf}  

\usepackage{amsmath,amsfonts} 
\usepackage{amssymb}  
\newcommand{\mrt}{\mathrm{T}}
\newcommand{\rmt}{\mathrm{T}}
\usepackage{algorithm}
\usepackage{algpseudocode}
\usepackage{graphicx}
\usepackage{color}
\usepackage{booktabs}
\usepackage{cite}

\title{\LARGE \bf
Robust Optimal Control With Binary Adjustable Uncertainties
}

\author{Yun Li$^{1}$, Neil Yorke-Smith$^2$ and Tamas Keviczky$^{1}$
\thanks{*The work was supported by the Brains4Buildings project under the Dutch grant programme for Mission-Driven Research, Development and Innovation (MOOI).}
\thanks{$^{1}$Yun Li and Tamas Keviczky are with Delft Center for Systems and Control, 
        Delft University of Technology,  Delft, the Netherlands.
        {\tt\small y.li-39@tudelft.nl, T.Keviczky@tudelft.nl}}%
\thanks{$^{2}$Neil Yorke-Smith is with STAR Lab, Delft University of Technology, the Netherlands. {\tt\small n.yorke-smith@tudelft.nl}
        {\tt\small }}%
}

\begin{document}

\maketitle
\thispagestyle{empty}
\pagestyle{empty}

\begin{abstract}
Robust Optimal Control (ROC) with adjustable uncertainties has proven to be effective in addressing critical challenges within modern energy networks, especially the reserve and provision problem. However, prior research on ROC with adjustable uncertainties has predominantly focused on the scenario of uncertainties modeled as continuous variables. In this paper, we explore ROC with binary adjustable uncertainties, where the uncertainties are modeled by binary decision variables, marking the first investigation of its kind.
To tackle this new challenge, firstly we introduce a metric designed to quantitatively measure the extent of binary adjustable uncertainties. Then, to balance computational tractability and adaptability, we restrict control policies to be affine functions with respect to uncertainties, and propose a general design framework for ROC with binary adjustable uncertainties. To address the inherent computational demands of the original ROC problem, especially in large-scale applications, we employ strong duality (SD) and big-M-based reformulations to create a scalable and computationally efficient Mixed-Integer Linear Programming (MILP) formulation. Numerical simulations are conducted to showcase the performance of our proposed approach, demonstrating its applicability and effectiveness in handling binary adjustable uncertainties within the context of modern energy networks.


\end{abstract}

\section{INTRODUCTION}\label{sec:1}
In recent years, the so-called reserve and provision problem has gained increasing attention owing to its ability to characterize the demand side management issues in modern energy networks. As an efficient approach for modeling the reserve and provision problem, a new optimal control framework called \textit{robust optimal control} (ROC) \textit{with adjustable uncertainties} was proposed \cite{bitlisliouglu2017robust,zhang2017robust,vrettos2016robust,nohadani2018optimization,bunning2022robust,lappas2018robust}. Unlike the conventional robust optimal control problem, where the uncertainty sets are fixed and are determined by exogenous factors, ROC with adjustable uncertainties considers the case where the scope of uncertainties is adjustable and is determined by decision makers. 

By allowing the decision maker to actively decide the admissible scope of uncertainty, the corresponding robust optimal control framework can be more versatile to include more practical cases. However, this versatility comes with more computational challenges. It is worth pointing out that existing research about ROC with adjustable uncertainties all focus on the case of continuous adjustable uncertainties. In the continuous case, uncertainty sets are typically represented as polyhedra or norm balls. As continuous variables yield an infinite number of uncertain scenarios within a given set, this leads to a semi-infinite optimization problem in the ROC formulation with adjustable uncertainties. To create a computationally tractable reformulation, strong duality theory is employed, reducing the need to satisfy universal constraints for all uncertain scenarios and instead focusing only on the worst-case scenario  \cite{bitlisliouglu2017robust,zhang2017robust,vrettos2016robust,nohadani2018optimization,bunning2022robust,lappas2018robust}.

Despite the success of ROC with continuous adjustable uncertainties in supporting modern energy networks, it should be noted that it fails to deal with the case where uncertainties can be binary. For example, in building climate control, some heating, ventilation and air conditioning (HVAC) devices only operate with two modes: on and off. If such devices are considered for providing demand response services, the existing ROC with continuous adjustable uncertainties fails to characterize this scenario. In addition, the existing design framework for ROC with continuous adjustable uncertainties cannot be directly generalized to deal with the binary case. On the one hand, polyhedra or norm balls used in the continuous case to describe the scope of uncertainties is not suitable for binary uncertainties. On the other hand, duality-based reformulation, which is built upon the assumption of continuous uncertainties, utilized in the continuous case for finding the worst uncertainty scenario cannot be directly applied to deal with the binary case. Motivated by this fact, in this paper, we investigate the problem of ROC with binary adjustable uncertainties. Our main contributions can be summarized as follows:
\begin{itemize}
    \item The problem of ROC with binary adjustable uncertainties is investigated for the first time. By introducing a new metric to evaluate the scope of binary uncertainties, we propose a general design formulation for ROC with binary adjustable uncertainties.
    \item Due to the high computational demand and poor scalability of the orignal formulation of ROC with binary adjustable uncertainties, we provide a computationally efficient and scalable alternative together with detailed derivation procedures. In addition, the probabilistic robustness of the optimal solution for our proposed approach is analyzed via Markov inequality.
    \item We consider a practical case of reserve and provision problem: utilizing the energy flexibility of buildings to provide demand response services and show how our proposed approach can be utilized to explore the energy flexibility potential. Simulation results are provided to demonstrate the effectiveness of the proposed approach. 
\end{itemize}
The remaining parts of this paper are organized as follows. Section \ref{sec:2} introduces the control problem and proposes a novel formulation of the ROC with binary adjustable uncertainties. For this original formulation, Section \ref{sec:3} gives a computationally efficient alternative with detailed derivation procedures and also analyzes the probabilistic robustness of the optimal solution. Simulation results are presented in Section \ref{sec:4}, and the conclusion follows in Section \ref{sec:5}.

\textbf{Notation}: Bold-face letters are used to denote the stacked time sequences of related signals; the operator $|\cdot|$ for a given set of indices denotes the cardinality of the set; subscript $i$ of a given vector/matrix represents the $i$-th element/row of the corresponding vector/matrix; $\mathbb{N}_{[0,a]}$ denotes the set of nonnegative integers less than equal to $a$; $\max$ or $\min$ for a given vector implies row-wise maximization/minimization. $\sum_kx_k$ denotes the summation of all terms $x_k$, $\mathbf{1}_N$ denotes the $N$-dimensional all 1 vector. 

\section{Problem Formulation}\label{sec:2}
In this section, we will introduce the problem of robust optimal control with binary adjustable uncertainties. The considered system is 
\begin{equation}\label{eq:sys}
    x(t+1) = Ax(t) + Br(t) + Du(t) + Ev(t)
\end{equation}
where the $x(t)\in\mathbb{R}^n$ are system states, $r(t)\in\mathbb{B}^m$ are reference control inputs, $u(t)\in\mathbb{R}^p$ and $v(t)\in\mathbb{B}^q$ are continuous and binary recourse control variables, respectively. For system \eqref{eq:sys}, we assume that the following linear state and input constraints should be satisfied
\begin{subequations}\label{eq:temp_cons}
\begin{align}
& G_xx(t) \leq g_x, \\
& G_rr(t) + G_uu(t) + G_v v(t) \leq g_r.
\end{align}
\end{subequations}

In our design framework, the reference control input is assumed to be predetermined but suffers from adjustable uncertainties, which is equivalent to the case that the system is subject to additive uncertainties. 

Within the context of the reserve and provision problem, system \eqref{eq:sys} can model the dynamics of several utilities in energy systems, such as building thermal dynamics or energy storage devices. For example, for building climate control, system \eqref{eq:sys} can model the thermal dynamics of buildings that draw power from both main power grid $r(t)$, which is usually determined according to the day-ahead electricity market, and local renewable energy sources (RES) ($u(t)$ and $v(t)$), which can be flexibly scheduled in real time.

Based on \eqref{eq:sys}, the predicted state evolution over $N$ time steps is
\begin{equation}
    \mathbf{x} = F_xx(0) + F_r \mathbf{r} + F_{u}\mathbf{u} + F_{v}\mathbf{v},
\end{equation}
where $\mathbf{x} = [x(1)^{\mrt},\cdots,x(N)^{\mrt}]^{\mrt}\in\mathbb{R}^{nN}$, $\mathbf{r} = [r(0)^{\mrt},\cdots,r(N-1)^{\mrt}]^{\mrt}\in\mathbb{B}^{mN}$, $\mathbf{u}= [u(0)^{\mrt},\cdots,u(N-1)^{\mrt}]\in\mathbb{R}^{pN}$, and $\mathbf{v} = [v(0)^{\mrt},\cdots,v(N-1)^\mrt]^\mrt\in\mathbb{B}^{qN}$, and $x^0$ is the initial state vector. The detailed format of the matrixes $F_x$, $F_r$, $F_u$ and $F_v$ can be found in \cite{yun23}.

Correspondingly, constraints \eqref{eq:temp_cons} within the prediction horizon can be compactly denoted as
\begin{subequations}\label{eq:constraint}
    \begin{align}
    &\mathbb{X}^N := \{\mathbf{x}|G_{\mathbf{x}}\mathbf{x} \leq g_{\mathbf{x}}\}, \\ 
    &\mathbb{U}^N := \{(\mathbf{u},\mathbf{v})|G_{\bf r}\mathbf{r} + G_{\mathbf{u}}\mathbf{u} + G_{\mathbf{v}}\mathbf{v} \leq g_{\bf r}\},
    \end{align}
\end{subequations}
where $G_{\bf x} = \text{diag}(G_x,\cdots,G_x)$, $G_{\bf r} = \text{diag}(G_r,\cdots,G_r)$, $G_{\bf u} = \text{diag}(G_u,\cdots,G_u)$, $G_{\bf v} = \text{diag}(G_v,\cdots,G_v)$, $g_{\bf x} = \mathbf{1}_N\otimes g_x$, and $g_{\bf r} = \mathbf{1}_N\otimes g_r$.

In the reserve and provision framework, the reference control signal suffers from unknown external manipulations to provide services for external entities, and the design objective is to optimally quantify how much uncertainty (also called flexibility) can be reserved while guaranteeing system constraints. For the case of continuous adjustable uncertainties, the scope of flexibility is denoted as adjustable polyhedrons or norm balls, and the volume of uncertainty sets are adopted as metrics to measure the scope of uncertainties\cite{zhang2017robust}. However, in the case of binary uncertainty, this definition of uncertainty set is not applicable because polyhedron/norm balls cannot accurately characterize uncertain binary variables, and their volumes also cannot truly reflect the scope of uncertainties.

In the following, we introduce a new metric to deal with binary adjustable uncertainties. The indices $k\in\mathbb{N}_{[0,mN-1]}$ for $\mathbf{r}_k$ within the prediction horizon are partitioned into two disjunctive groups $\mathcal{C}$ and $\mathcal{U}$. For all $k\in\mathcal{U}$, the corresponding reference inputs $\mathbf{r}_k$ are allowed to be adjusted flexibly. For all $k\in\mathcal{C}$, the reference inputs $\mathbf{r}_k$ are not influenced by uncertainties and are fixed to their nominal values $\mathbf{\bar r}_k$. 

\begin{figure}
    \centering
    \includegraphics[width = 0.45\textwidth]{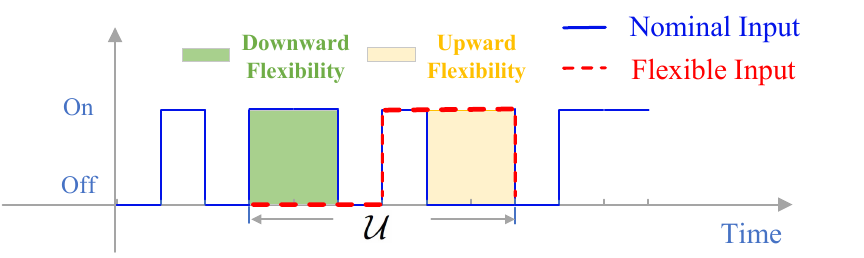}
    \caption{Schemetic diagram of flexible reference input signal $\bf r$.}
    \label{fig:flexi}
\end{figure}

However, due to system constraints, not all reference inputs $\mathbf{r}_k(k\in\mathcal{U})$ can be flexibly adjusted without violating \eqref{eq:temp_cons}. Here, we define $\mathcal{S}\subseteq\mathcal{U}$ as the index set in which the corresponding $\mathbf{r}_k$ ($k\in\mathcal{S}$) is adjusted from $\bar{\bf r}_k$ to a new value, denoted as $\tilde{\bf r}_k$ with $\tilde{\bf r}_k \neq \bar{\bf r}_k$, without violating system constraints. Based on the above definition, the values of $\mathbf{r}_k$ within the prediction horizon can be represented as
\begin{equation}\label{eq:uncertainty}
    \mathbf{r}_k = 
    \begin{cases}
    \tilde{\mathbf{r}}_k, & k\in\mathcal{S} \\
    \bar{\mathbf{r}}_k, & k\in\mathcal{C}\cup \mathcal{U}\setminus\mathcal{S}
    \end{cases}
\end{equation}
A schematic diagram of flexible reference input signals with $m = 1$ is given in Fig. \ref{fig:flexi}. Similarly to the case of continuous adjustable uncertainty,  $\bar{\bf r}_k = 0$ and $\tilde{\bf r}_k = 1$ means up-ward flexibility, and $\bar{\bf r}_k=1$ and $\tilde{\bf r}_k = 0$ means down-ward flexibility.

In contrast to the continuous case, where the volume of the uncertainty set is adopted to measure the scope of reserved flexibility, according to the definition \eqref{eq:uncertainty}, a new variable $\Gamma$ is defined as the largest cardinality of $\mathcal{S}$ that can be freely selected among $\mathcal{U}$, to measure the scope of the reserved uncertainty/flexibility.
An admissible $\Gamma$ implies that there exist feasible recourse inputs $(\mathbf{u,v})$ to ensure system constraint satisfaction for any possible reference input adjustment as long as $|\mathcal{S}|\leq \Gamma$. 
Clearly, a larger value of $\Gamma$ implies that the controlled system can reserve more flexibility for providing services. 

For the problem of ROC with adjustable uncertainties, a general control objective is to balance the minimization of operational cost and the maximization of the scope of reserved flexibility, while guaranteeing constraint satisfaction under all possible scenarios of uncertainties/flexibilities \cite{bitlisliouglu2017robust,zhang2017robust}. Accordingly, the design objective for our considered problem takes the form
\begin{subequations}\label{eq:formulation}
\begin{align}
    \min_{\mathbf{u,v},\Gamma}\ & \big\{\max_{\mathcal{S}} \{J(\mathbf{x},\mathbf{u},\mathbf{v})\} - \lambda\Gamma\big\} \label{eq:obj}\\
    \text{s.t. } & \mathbf{x} = F_xx^0 + F_r\mathbf{r} + F_u\mathbf{u} +F_v\mathbf{v}\\
    & \mathbf{x} \in \mathbb{X}^N,\ (\mathbf{u},\mathbf{v}) \in \mathbb{U}^N, \label{eq:7b} \\
    & \mathbf{r}_k = \bar{\bf r}_k,\ \forall k \in \mathcal{C}\cup \mathcal{U}\setminus \mathcal{S},\\
    & \mathbf{r}_k = \tilde{\mathbf{r}}_k,\ \forall k \in\mathcal{S},\\
    & \forall \mathcal{S}: |\mathcal{S}| \leq \Gamma,
\end{align}
\end{subequations}
where $J(\mathbf{x},\mathbf{u},\mathbf{v}):= \sum_{k=0}^Nl_k(\mathbf{x},\mathbf{u},\mathbf{v})$ is the summation of linear stage cost functions $l_k(\mathbf{x},\mathbf{u},\mathbf{v})$, $\lambda>0$ is a user-selected weighting parameter.  

The objective function \eqref{eq:obj} aims at minimizing the worst-case operational cost given $\Gamma$, namely $\max_{\mathcal{S}}J(\mathbf{x,u,v})$, w.r.t. the input uncertainties determined by $\mathcal{S}$ and in the meanwhile maximizing the scope of flexibility $\Gamma$. If the worst-case operational cost function $\max_{\mathcal{S}}J(\mathbf{x,u,v})$ is removed, optimization problem \eqref{eq:formulation} is to find out the largest feasible $\Gamma$. Namely, the largest cardinality of $\mathcal{S}$ that can be freely selected without violating system constraints. For example, in Fig. \ref{fig:flexi}, four inputs among $\{\mathbf{r}_k:\forall k\in\mathcal{U}\}$ are adjusted, if the optimal $\Gamma^*$ computed in \eqref{eq:formulation} satisfies $\Gamma^*\geq 4$, we are still capable of ensuring system constraints for this scenario of flexible inputs.

To ensure the feasibility of \eqref{eq:formulation}, it is assumed that for the predefined reference input $\bar{\mathbf{r}}$, there exists at least one input sequence $(\mathbf{u,v})$ such that the input constraints \eqref{eq:7b} are satisfied.

\textit{Remark 1}: We refer to the optimal control problem \eqref{eq:formulation}, similarly to its continuous counterparts: ROC with continuous adjustable uncertainties  \cite{bitlisliouglu2017robust,zhang2017robust}, as \textit{ROC with binary adjustable uncertainties}. The main difference between ROC with adjustable uncertainties and conventional ROC problem lies in the uncertainty set. For conventional ROC problems, the uncertainty sets are predetermined and fixed. But for the ROC with adjustable uncertainties, the shape/size of uncertainty sets are not fixed but will be optimally determined by decision variables, such as $\Gamma$ in \eqref{eq:formulation}. 

\textit{Remark 2}: We highlight that the problem formulation proposed in \eqref{eq:formulation} can be applied to several important issues in modern energy systems. For example, similarly to \cite{vrettos2016robust,bunning2022robust}, formulation \eqref{eq:formulation} can model the reserve and provision problem of buildings with on-off types of HVAC devices for providing demand response services. Besides, as shown in \cite{zhao2013vulnerability}, it is also possible to extend the formulation \eqref{eq:formulation} to analyze and design resilient control strategies for power grid networks to mitigate faults or cyberattacks.

\section{Robust Optimal Control Design and Robustness Analysis}\label{sec:3}
This section is devoted to solving the problem of ROC with binary adjustable uncertainties formulated in \eqref{eq:formulation}, and to analyzing the robustness of the derived solution.
\subsection{Robust Optimal Control Design}
Considering the system dynamics \eqref{eq:sys}, state and input constraints \eqref{eq:constraint}, and the linear property of the objective function, the optimization problem in \eqref{eq:formulation} can be rewritten in the following compacted form
\begin{subequations}\label{eq:formulation_compact}
    \begin{align}
        \min_{\theta,\Gamma,\mathbf{u,v}}\ & \theta - \lambda \Gamma \\
        \text{s.t. } & \mathbf{O}\mathbf{r} + \mathbf{P}\mathbf{u} + \mathbf{Q}\mathbf{v} \leq \mathbf{h}, \label{eq:cons_orig}\\
        & \mathbf{r}_k = \bar{\bf r}_k,\ \forall k \in \mathcal{C} \cup \mathcal{U}\setminus \mathcal{S}, \\
        & \mathbf{r}_k = \tilde{\bf r}_k,\ \forall k \in \mathcal{S},\\
        &\forall\ \mathcal{S}:\ |\mathcal{S}| \leq \Gamma,\label{eq:uni_quanti}
    \end{align}
\end{subequations}
where matrixes $\mathbf{O}$, $\mathbf{P}$, $\mathbf{Q}$ and $\mathbf{h}$ are constructed according to system dynamics, constraints and problem data, $\theta$ is introduced for the epigraph reformulation of the worst-case operational cost in \eqref{eq:formulation} as robust constraints $J(\mathbf{x,u,v}) \leq \theta$ $\forall \mathcal{S}:|\mathcal{S}|\leq \Gamma$, which are compactly included in \eqref{eq:cons_orig}. Similar reformulation can be found in \cite{zhang2017robust}.

Unlike the continuous case, in which the original formulation of ROC with adjustable uncertainties is semi-infinite and is computationally intractable, the optimization problem \eqref{eq:formulation_compact} can be solved directly without further reformulation since it only entails a finite number of uncertainties $\tilde{\bf r}_k$ and $\mathcal{S}$ given a fixed $\Gamma$. By trying out all possible $\Gamma$ and considering all corresponding uncertain scenarios, the optimal solution can be found by exhaustive search. However, it should be noted that given a fixed $\Gamma$, the number of constraints that need to be considered in \eqref{eq:formulation_compact} is proportional to $\sum_{i=0}^\Gamma\binom{|\mathcal{U}|}{i}\cdot N$, which can result in a large scale MILP that is computationally demanding. In the following, we will propose a novel reformulation for \eqref{eq:formulation_compact} such that the resulting problem is scalable and computationally efficient.

For the recourse control inputs $\bf u$ and $\bf v$, we will design closed-loop control policies instead of open-loop control actions since it has been demonstrated in many works that closed-loop control policies can tolerate more uncertainties than open-loop control actions \cite{ben2004adjustable,bertsimas2022,bertsimas2018binary}. Because considering arbitrary control policies can make the problem computationally intractable, we consider the following affine control policies to balance robustness, optimality, and computational effort for our proposed scheme
\begin{subequations}\label{eq:control_policy}
\begin{align}
    &\mathbf{u}_k = \sum_{i\in\mathcal{U}}M_{ki}\mathbf{r}_i + \eta_k,\\
    &\mathbf{v}_k = \sum_{i\in\mathcal{U}}L_{ki}\mathbf{r}_i + \epsilon_k,\label{eq:ploicy_v}
\end{align}
\end{subequations}
where $M_{ki}\in\mathbb{R}$, $\eta_k\in\mathbb{R}$, $L_{ki}\in\mathbb{Z}$ and $\epsilon_k\in\mathbb{Z}$ are control policy parameters to be optimized. 
Since the exact value of the flexible inputs $\tilde{\bf r}_k (k\in\mathcal{U})$ is not determined when solving \eqref{eq:formulation_compact}, the control policy \eqref{eq:control_policy} will adapt inputs $\mathbf{u}_k$ and $\mathbf{v}_k$ once ${\mathbf{r}}_i (i\in\mathcal{U})$ are revealed, which can increase the robustness of the controlled system so that more flexibility can be reserved. Note that binary restriction for $\mathbf{v}_k$ can be imposed by a linear constraint $0\leq \mathbf{v}_k \leq 1$ due to the fact that the control policy $\eqref{eq:ploicy_v}$ will only generate integer outputs.

\textbf{Theorem 1}: Considering the system dynamics \eqref{eq:sys}, the binary adjustable input uncertainties 
\eqref{eq:uncertainty} and the control policies \eqref{eq:control_policy}, the robust optimal control problem in \eqref{eq:formulation_compact} can be reformulated as the following MILP problem
\begin{subequations}\label{eq:final}
    \begin{align}   \min_{\substack{\theta,\delta_k,\mu_{ij},\phi_{ij},y_{ij}\beta_{ij},\\M_{kj},L_{kj},\eta_k,\epsilon_k}}\ & \theta - \lambda \sum_{k\in\mathcal{U}}\delta_k \\
        \text{s.t. }& \sum_{j\in\mathcal{U}}\mu_{ij} + \sum_{j\in\mathcal{U}}\beta_{ij} + \phi_i +\sum_{k}\mathbf{P}_{ik}\eta_k + \notag\\
        &\qquad \sum_{k}\mathbf{Q}_{ik}\epsilon_k + \sum_{j\in \mathcal{C}}\mathbf{O}_{ij}\bar{\mathbf{r}}_j  \leq \mathbf{h}_i, \label{eq:constraint_compact}\\
        & \mu_{ij} + \pi_i \geq \frac{1}{2}\bigg( \mathbf{O}_{ij} + \sum_{k}\mathbf{P}_{ik}M_{kj} + \notag\\
        & \qquad \sum_k \mathbf{Q}_{ik}L_{kj}+ y_{ij}\bigg) + \bigg(\mathbf{O}_{ij} + \notag\\
        &\qquad \sum_{k}\mathbf{P}_{ik}M_{kj} + \sum_k\mathbf{Q}_{ik}L_{kj}\bigg)\bar{\bf r}_j,\\
        & -y_{ij} \leq \mathbf{O}_{ij} + \sum_k {\bf P}_{ik}M_{kj} + \notag\\
        &\qquad \sum_k\mathbf{Q}_{ik}L_{kj}\leq y_{ij},\\
        & \phi_{i} \geq \sum_{j\in\mathcal{U}}\bigg(\mathbf{O}_{ij}+\sum_k\mathbf{P}_{ik}M_{kj}+\notag\\
        & \qquad \sum_k\mathbf{Q}_{ik}L_{kj}\bigg)\cdot\mathbf{\bar{r}}_j,\\
        & 0 \leq \beta_{ij} \leq M\delta_j, \\
        & 0 \leq \pi_i - \beta_{ij} \leq M(1-\delta_j), \\
        & \delta_j\in\mathbb{B},\\
        & \mu_{ij}\geq 0,\ \pi_i\geq 0,\ y_{ij}\geq 0,\\
        & i\in\mathcal{I},\ j\in\mathcal{U},
    \end{align}
\end{subequations}
where $\mathcal{I}$ is the set of row indices of constraints \eqref{eq:cons_orig}, and the optimal solution $\delta_k^*$ satisfies $\sum_{j\in\mathcal{U}}\delta_j^* = \Gamma^*$.

\textit{Proof}: Given a fixed $\Gamma$, system constraint \eqref{eq:cons_orig} has to be satisfied for all possible $\tilde{\mathbf{r}}_k$ and $\mathcal{S}$ such that $k\in\mathcal{S}$ and $|\mathcal{S}|\leq \Gamma$. To ensure constraint satisfaction for all possible uncertainties, which entails the order of $\sum_{i=0}^{\Gamma}\binom{|\mathcal{U}|}{i}$ uncertainty scenarios, we alternatively design a control scheme to guarantee constraint satisfaction under the worst case scenario of uncertainty, which can be represented as
\begin{equation}\label{eq:worst_case_cons}
         \max_{\mathcal{S}} \big\{\mathbf{Or+Pu+Qv}:|\mathcal{S}|\leq \Gamma, \mathbf{r}_j \text{ satisfies \eqref{eq:uncertainty}}\big\} \leq \mathbf{h},
\end{equation}
where the maximization and inequality in \eqref{eq:worst_case_cons} is computed and hold row-wise, respectively.
The $i$-th row of constraints \eqref{eq:worst_case_cons}, combining the control policies \eqref{eq:control_policy} and the reference input pattern \eqref{eq:uncertainty} can be expressed as
\begin{equation*}
    \begin{split}
        \max_{\mathcal{S}:|\mathcal{S}|\leq\Gamma} \ &\bigg\{ \mathbf{O}_i^\rmt\mathbf{r} + \sum_k\mathbf{P}_{ik}\bigg(\sum_{j\in\mathcal{U}}M_{kj}\mathbf{r}_j + \eta_k\bigg) + \\
        &\sum_k\mathbf{Q}_{ik}\bigg(\sum_{j\in\mathcal{U}}L_{kj}\mathbf{r}_j + \epsilon_k\bigg)\bigg\} \leq \mathbf{h}_i, 
    \end{split}
\end{equation*}
which can be further reformulated as
\begin{eqnarray}\label{eq:constraint_worst}
&& \max_{\mathcal{S}:|\mathcal{S}|\leq \Gamma} \bigg\{  \sum_{j\in\mathcal{S}}\bigg(\mathbf{O}_{ij} + \sum_k\mathbf{P}_{ik}M_{kj} +\sum_{k}\mathbf{Q}_{ik}L_{kj}\bigg)\tilde{\bf r}_j \notag\\
        &&\qquad+ \sum_{j\in \mathcal{U}\setminus \mathcal{S}} \bigg(\mathbf{O}_{ij} +\sum_k \mathbf{P}_{ik}M_{kj} + \sum_k\mathbf{Q}_{ik}L_{kj}\bigg)\bar{\bf r}_j \bigg\} \notag\\
        &&\qquad + \sum_{k}\mathbf{P}_{ik}\eta_k + \sum_{k}\mathbf{Q}_{ik}\epsilon_k  + \sum_{j\in\mathcal{C}}\mathbf{O}_{ij}\bar{\mathbf{r}}_j  \leq \mathbf{h}_i.    
\end{eqnarray}
As explained in \cite{ramirez2022robust}, it can be verified that 
\begin{equation}
    \begin{split}
        &\bigg(\mathbf{O}_{ij} + \sum_{k}\mathbf{P}_{ik}M_{kj} + \sum_{k}\mathbf{Q}_{ik}L_{kj}\bigg)\tilde{\bf r}_j \leq \\
         & \qquad      
        \frac{1}{2}\bigg(\mathbf{O}_{ij} + \sum_{k}\mathbf{P}_{ik}M_{kj} + \sum_{k}\mathbf{Q}_{ik}L_{kj}
        + \big|\mathbf{O}_{ij} + \\
        &\qquad \sum_{k}\mathbf{P}_{ik}M_{kj} + \sum_k\mathbf{Q}_{ik}L_{kj}\big|\bigg).
    \end{split}
\end{equation}
Accordingly, by following a similar line as in \cite{bertsimas2004price} and introducing additional decision variables $z_j\in\mathbb{R}\ (j\in\mathcal{U})$, the maximization problem in \eqref{eq:constraint_worst} is relaxed as
\begin{subequations}\label{eq:maximization_z}
    \begin{align}  
    \max_{z_{j}}\ &\sum_{j\in\mathcal{U}}\frac{z_j}{2}\cdot \bigg( \mathbf{O}_{ij}+\sum_{k}\mathbf{P}_{ik}M_{kj}  + \sum_k \mathbf{Q}_{ik}L_{kj} + \notag\\
    &\quad  |\mathbf{O}_{ij}  +\sum_{k}\mathbf{P}_{ik}M_{kj} + \sum_k \mathbf{Q}_{ik}L_{kj}| \bigg) + \notag \\
    &\quad (1-z_j)\cdot\bigg( \mathbf{O}_{ij} +\sum_{k}\mathbf{P}_{ik}M_{kj} + \sum_{k}\mathbf{Q}_{ik}L_{kj}\bigg) \bar{\bf r}_j\\
    \text{s.t. }& 0 \leq z_j \leq 1,\quad \sum_{j\in\mathcal{U}}z_j \leq \Gamma.
    \end{align}
\end{subequations}
Since \eqref{eq:maximization_z} is a linear program (LP) w.r.t. $z_j$, according to strong duality theory of LP, see \cite{boyd2004convex}, the objective value of \eqref{eq:maximization_z} coincides with that of the following dual problem
\begin{subequations}\label{eq:dual_z}
    \begin{align}
        \min_{\mu_{ij},\pi_i,\phi_i}\ & \sum_{j\in\mathcal{U}}\mu_{ij} + \Gamma\pi_i + \phi_{i} \\
        \text{s.t. }& \mu_{ij} + \pi_i \geq \frac{1}{2}\bigg(\mathbf{O}_{ij} +\sum_{k}\mathbf{P}_{ik}M_{kj}  + \sum_k\mathbf{Q}_{ik}L_{kj} \notag\\
        &\quad + |\mathbf{O}_{ij} +\sum_{k}\mathbf{P}_{ik}M_{kj} + \sum_k\mathbf{Q}_{ik}L_{kj} | \bigg) - \notag\\
        &
        \quad \bigg(\mathbf{O}_{ij} +\sum_{k}\mathbf{P}_{ik}M_{kj} + \sum_k\mathbf{Q}_{ik}L_{kj}\bigg)\bar{\bf r}_j, \\
        & \phi_{i} \geq \sum_{j\in\mathcal{U}}\bigg(\mathbf{O}_{ij}+\sum_k\mathbf{P}_{ik}M_{kj}+\sum_k\mathbf{Q}_{ik}L_{kj}\bigg)\mathbf{\bar r}_j,\\
        & \mu_{ij}\geq 0,\ \pi_i\geq 0,\ \forall j\in\mathcal{U},
    \end{align}
\end{subequations}
where $\mu_{ij}$ and $\pi_{i}$ are Lagrange multipliers.

Optimization problem \eqref{eq:dual_z} is nonlinear because it contains the absolute function, which will incur increased computational effort. By introducing auxiliary decision variables $y_{ij}\geq 0$, optimization problem \eqref{eq:dual_z} can be further relaxed as the following linear optimization problem
\begin{subequations}\label{eq:reformulated}
    \begin{align}
            \min_{\mu_{ij},\pi_i,\phi_i}\ & \sum_{j\in\mathcal{U}}\mu_{ij} + \Gamma\pi_i + \phi_{i} \\
        \text{s.t. }& \mu_{ij} + \pi_i \geq \frac{\mathbf{O}_{ij} +\sum_{k}\mathbf{P}_{ik}M_{kj} + \sum_{k}\mathbf{Q}_{ik}L_{kj}+ y_{ij}}{2}  \notag \\
        &\quad  -\bigg(\mathbf{O}_{ij} +\sum_{k}\mathbf{P}_{ik}M_{kj} + \sum_k\mathbf{Q}_{ik}L_{kj}\bigg)\cdot   \bar{\bf r}_j, \\
        & -y_{ij} \leq \mathbf{O}_{ij} +\sum_{k}\mathbf{P}_{ik}M_{kj} + \sum_k\mathbf{Q}_{ik}L_{kj}\leq y_{ij},\\
        & \phi_{i} \geq \sum_{j\in\mathcal{U}}\bigg(\mathbf{O}_{ij}+\sum_k\mathbf{P}_{ik}M_{kj}+\sum_k\mathbf{Q}_{ik}L_{kj}\bigg)\cdot\mathbf{\bar{r}}_j, \\
        & \mu_{ij}\geq 0,\ \pi_i\geq 0,\ y_{ij}\geq 0,\ \forall j\in\mathcal{U}.  
    \end{align}
\end{subequations}
For the inequality \eqref{eq:constraint_worst}, the maximization problem can be relaxed by a feasible solution of the above LP problem to yield the following alternative constraints
\begin{subequations}\label{eq:final_relaxed_cons}
    \begin{align}
        &\sum_{j\in\mathcal{U}}\mu_{ij}+\Gamma\pi_i + \phi_i+ \sum_k\mathbf{P}_{ik}\eta_k + \sum_{k}\mathbf{Q}_{ik}\epsilon_k +\notag\\
        &\qquad \sum_{j\in\mathcal{C}}\mathbf{O}_{ij}\bar{\bf r}_j\leq \mathbf{h}_i,\label{eq:17a}\\
        & \mu_{ij} + \pi_i \geq \frac{1}{2}\bigg(\mathbf{O}_{ij} +\sum_{k}\mathbf{P}_{ik}M_{kj} + \sum_k\mathbf{Q}_{ik}L_{kj}+ y_{ij}\bigg) \notag \\
        &\quad  - \bigg(\mathbf{O}_{ij} +\sum_{k}\mathbf{P}_{ik}M_{kj} + \sum_k\mathbf{Q}_{ik}L_{kj} \bigg)\cdot\bar{\bf r}_j, \\
        & -y_{ij} \leq \mathbf{O}_{ij} +\sum_{k}\mathbf{P}_{ik}M_{kj} + \sum_k\mathbf{Q}_{ik}L_{kj}\leq y_{ij},\\
        & \phi_{i} \geq \sum_{j\in\mathcal{U}}\bigg(\mathbf{O}_{ij}+\sum_k\mathbf{P}_{ik}M_{kj}+\sum_k\mathbf{Q}_{ik}L_{kj}\bigg)\cdot\mathbf{\bar{r}}_j,\\
        & \mu_{ij}\geq 0,\ \pi_i\geq 0,\ y_{ij}\geq 0,\ \forall j\in\mathcal{U}. 
    \end{align}
\end{subequations}
Furthermore, introducing $\delta_j\in\mathbb{B}\ (j\in\mathcal{U})$ and $\beta_{ij} \in\mathbb{R}\ (i\in\mathcal{I},j\in\mathcal{U})$ and defining $\Gamma := \sum_{j\in\mathcal{U}}\delta_j$, the term $\Gamma\pi_i$ in \eqref{eq:17a} can be further relaxed via big-M formulation as
\begin{subequations}\label{eq:bilinear_reformu}
    \begin{align}
        &\Gamma\pi_i = \sum_{j\in\mathcal{U}} \beta_{ij}, \\
        &0 \leq \beta_{ij} \leq M\delta_j, \\
        &0\leq \pi_i - \beta_{ij} \leq M(1-\delta_j), \\
        &\delta_j\in\mathbb{B},
    \end{align}
\end{subequations}
where $M > 0$ is a sufficiently large constant. Finally, constraints \eqref{eq:cons_orig} -- \eqref{eq:uni_quanti} can be replaced by \eqref{eq:final_relaxed_cons} and \eqref{eq:bilinear_reformu}, which result in the optimization problem \eqref{eq:final}. This completes the proof. \hfill $\square$

\textit{Remark 3}: For the original formulation in \eqref{eq:formulation_compact}, given a fixed $\Gamma$, in order to guarantee constraint satisfaction robustly, the number of constraints considered is proportional to $\sum_{i=0}^\Gamma\binom{|\mathcal{U}|}{i}\cdot N$. In contrast, the number of constraints in the reformulated optimization problem in \eqref{eq:final} is only proportional to $|\mathcal{U}|\cdot N$, which is more favourable when $\sum_{i=0}^\Gamma\binom{|\mathcal{U}|}{i}\gg |\mathcal{U}|$. In addition, the constraint \eqref{eq:17a} has nonconvex term $\Gamma\pi_i$ since $\Gamma\in\mathbb{N}_{[0,|\mathcal{U}|]}$ is a decision variable. While this nonconvexity in \eqref{eq:17a} can be dealt with by several solvers, e.g. {\tt Gurobi}, with the big-M based reformualation in \eqref{eq:bilinear_reformu} the final optimization problem \eqref{eq:final}, which is a MILP, can be solved more efficiently by more off-the-shelf solvers, such as {\tt Gurobi}, {\tt CPLEX}, and {\tt GLPK}.

\subsection{Robustness of the Optimal Solution}
The obtained optimal control strategy, namely $(M^*_{kj},\eta^*_k,L_{kj}^*,\epsilon_k^*)$, is able to guarantee constraint satisfaction if no more than $\Gamma$ of the flexible reference inputs ${\mathbf{r}}_j\ (j\in\mathcal{U})$ are changed. In practice, however, it is possible that in some cases more than $\Gamma$ number of reference inputs $\mathbf{r}_k$ are subjected to uncertainties, e.g., in case of device faults. As a result, it is of interest and importance to analyze the robustness of the derived solution for such cases. In the following, we will investigate the possibility of constraint violation with the derived optimal solution when more than $\Gamma$ number of ${\mathbf{r}}_j\ (j\in\mathcal{U})$ might be changed.

In our analysis, $\mathbf{r}_j\ (j\in\mathcal{U})$ are regarded as random variables (r.v.). Consequently, the left-hand side of constraint \eqref{eq:cons_orig} are also r.v. For the $i$-th row of the constraint \eqref{eq:cons_orig}, its probability of violation is
\begin{multline}\label{eq:violatioin_p}
       \mathbb{P}_{\text{vio}} = \text{Pr}\big(\mathbf{O}_i^\rmt \mathbf{r} + \sum_{j\in\mathcal{U}}\mathbf{r}_j\sum_k\mathbf{P}_{ik}M^*_{kj} + \sum_{j\in\mathcal{U}}\mathbf{r}_j\sum_{k}\mathbf{Q}_{ik}L^*_{kj}+ \\
   \sum_k\mathbf{P}_{ik}\eta^*_k + \sum_k\mathbf{Q}_{ik}\epsilon^*_k \geq \mathbf{h}_i\big).
\end{multline}
For notational brevity, we define
\begin{subequations}\label{eq:ab_def}
\begin{align}
    & a_{ij}:= \mathbf{O}_{ij} + \sum_{k}\mathbf{P}_{ik}M_{kj}^* +\sum_k\mathbf{Q}_{ik}L^*_{kj},\\
    & b_i:= \mathbf{h}_i - \sum_{j\in\mathcal{C}}\mathbf{O}_{ij}{\bf r}_j - \sum_k\mathbf{P}_{ik}\eta_k^* - \sum_k\mathbf{Q}_{ik}\epsilon_k^*.
\end{align}
\end{subequations}

\textbf{Proposition 1}: Assuming that $\mathbf{r}_j\in\mathbb{B}\ (j\in\mathcal{U})$ are independent random variables, then the probability of violation for the $i$-th constraint \eqref{eq:violatioin_p} satisfies
\begin{equation}\label{eq:probability_violation}
    \mathbb{P}_{\text{vio}} \leq \frac{\prod_{j\in\mathcal{U}}\mathbb{E}\big(\exp(a_{ij}{\bf r}_j)\big)}{\exp(b_i)}.
\end{equation}
    
\textit{Proof}: Based on the definitions of $a_{ij}$ and $b_j$ in \eqref{eq:ab_def}, the probability of violation \eqref{eq:violatioin_p} can be rewritten as
\begin{subequations}
\begin{align}
    \mathbb{P}_{\text{vio}} &= \text{Pr} \bigg[ \sum_{j\in\mathcal{U}} a_{ij}\mathbf{r}_j  \geq b_i\bigg]\ \\
    & = \text{Pr}\left\{\exp\big(\sum_{j\in\mathcal{U}}a_{ij}\mathbf{r}_j\big) \geq \exp (b_i)\right\}.
    \end{align}
\end{subequations}
Then, it follows from the Markov inequality \cite{jacod2004probability} and the independence property of ${\bf r}_j$ that 
\begin{equation}
    \mathbb{P}_{\text{vio}} \leq \frac{\prod_{j\in\mathcal{U}}\mathbb{E}\big(\exp(a_{ij}{\bf r}_j)\big)}{\exp(b_i)}.
\end{equation}
This completes the proof. \hfill $\square$

It should be noted that the bound of the probability of violation in \eqref{eq:probability_violation} relies on system information in $a_{ij}$ and $b_i$. In the following, we give a probability bound that does not rely on system information and is easy to compute.

\textbf{Proposition 2}: Assuming that ${\bf r}_j\ (j\in\mathcal{U})$ are independent random variables with $\text{Pr}({\bf r}_j \neq \bar{\bf r}_j) = \epsilon_j$, Then, we have $\mathbb{P}_{\text{vio}} \leq \frac{\sum_{j\in\mathcal{U}}\epsilon_j}{\Gamma}$.

\textit{Proof}: Define random variables $z_j$ as the indicator functions of the events ${\bf r}_j \neq \bar{\bf r}_j\ (j\in\mathcal{U})$. Then, it is readily concluded that $\mathbb{P}_{\text{vio}} \leq \text{Pr}(\sum_{j\in\mathcal{U}}z_j \geq \Gamma)$. By applying Markov inequality and the independence of ${\bf r}_j$, it yields
\begin{subequations}
    \begin{align}
    \mathbb{P}_{\text{vio}}\ &  \leq \frac{\mathbb{E}(\sum_{j\in\mathcal{U}}z_j)}{\Gamma}\\
    & = \frac{\sum_{j\in\mathcal{U}}\mathbb{E}(z_j)}{\Gamma} = \frac{\sum_{j\in\mathcal{U}} \epsilon_j}{\Gamma}.
    \end{align}
\end{subequations}
This completes the proof.\hfill$\square$

While the bound of probability violation in Proposition 2 is independent of system information and is easy to compute, it is very conservative when $\Gamma$ is small or $\sum_{j\in\mathcal{U}}\epsilon_j$ is large.

\section{Simulation Results}\label{sec:4}

\begin{figure}[htb]
    \centering
    \includegraphics[width = 0.75\linewidth]{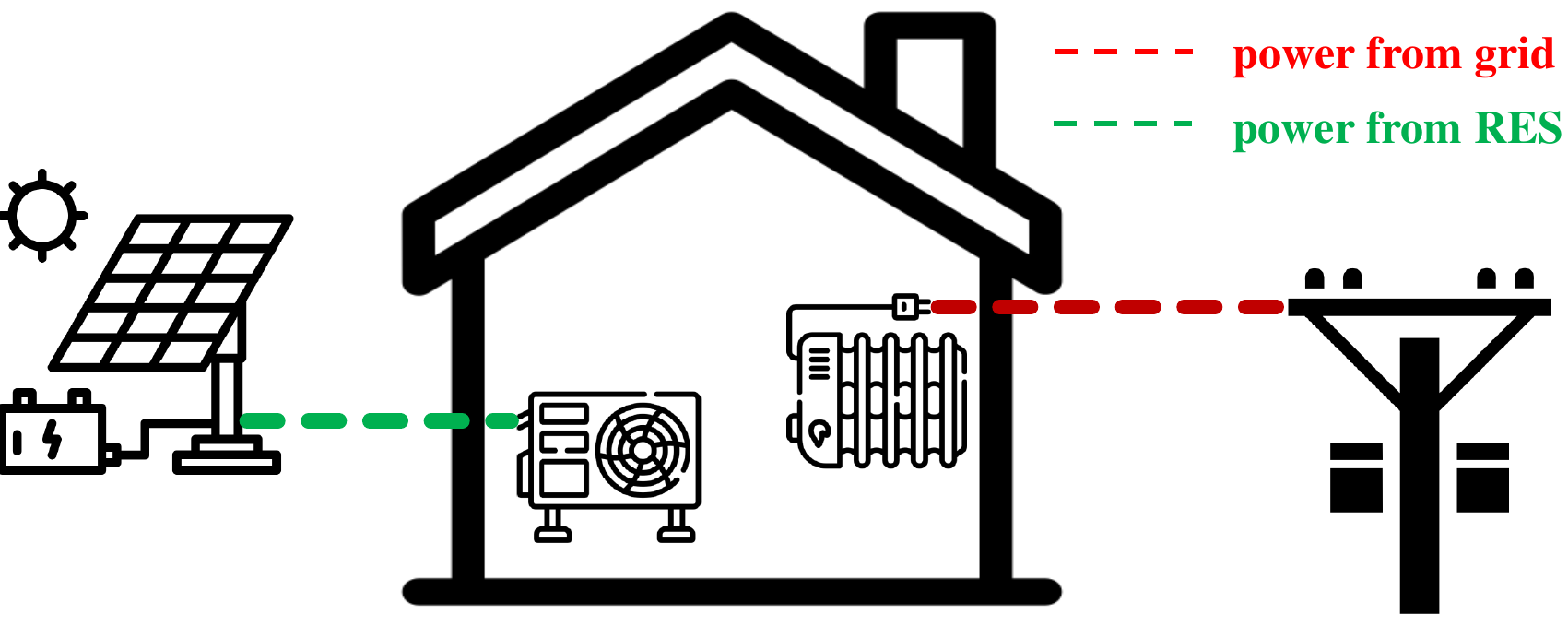}
    \caption{Diagram of energy consumption options.}
    \label{fig:building_structure}
\end{figure}
The effectiveness of our proposed approach is illustrated via a numerical case study about the reserve and provision problem for a smart building system. A schematic diagram of the considered building system is depicted in Fig. \ref{fig:building_structure}.
In our simulation, it is assumed that the thermal energy of the building is provided by three heat pumps (HP) that draw electricity from two sectors: power grid and local renewable energy sources (a solar panel). The HP powered by the grid has only two operation modes: on and off, which is denoted as $\bf r$, and the operation status is predetermined according to day-ahead electricity price. For the remaining two HPs drawing power from local RES, their control signals ($\bf u, v$) are available for adjusting in real-time. One HP has only on-off operational modes, and the other can operate continuously, e.g., PWM based operation. A solar panel is equipped to provide RES. 

As ubiquitous storage devices of energy flexibility, buildings are utilized to provide demand response (DR) services to the power grid. During demand response period, the building management system is asked to adjust its electricity consumption, namely the operational pattern of $\bf r$, according to the requests of the grid operator. 

In our simulation, the sampling period is selected as 30 min and the prediction horizon is set as $N=48$, i.e., one day. The starting time instant of flexibility assessment period $\mathcal{U}$ is selected as $22$ (11am). A solar panel with a maximal 2000W electricity output is included to provide RES. The initial indoor temperature is selected as $x(0) =21 ^{\circ}$C, and the indoor temperature bound is $[20^\circ\text{C},24^\circ \text{C}]$.

The objective for the optimization problem \eqref{eq:formulation_compact} is selected to maximize $\Gamma$. Namely, we want to assess the largest flexibility potential of the considered building system within the period of $\mathcal{U}$. Three schemes are adopted in our simulation:
\begin{itemize}
    \item Scheme 1: our proposed formulation \eqref{eq:final}.
    \item Scheme 2: original formulation \eqref{eq:formulation_compact} solved via exhaustive search.
    \item Scheme 3: our proposed formulation \eqref{eq:final} but only considering open-loop control actions ($M_{ki} = 0$, $L_{ki} = 0$ in \eqref{eq:control_policy}).
\end{itemize}
All simulations are carried out using Gurobi 9.5.1 \cite{gurobi} on an Intel Xeon W-2223 CPU at 3.60GHz with 16G RAM. Optimization problems are modeled via the Python package Pyomo \cite{bynum2021pyomo}. 


\begin{figure}
    \centering
    \includegraphics[width = 0.95\linewidth]{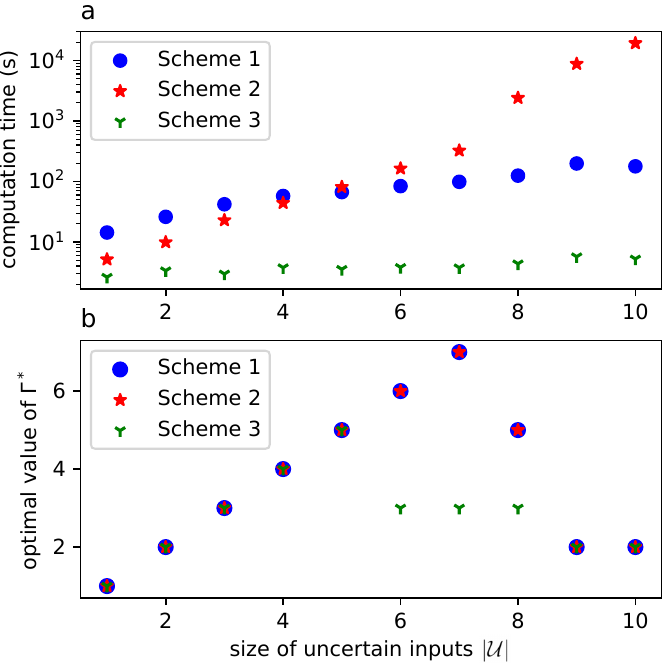}
    \caption{Computation time and optimal $\Gamma^*$ for all schemes.}
    \label{fig:time_value}
\end{figure}

\begin{figure}
    \centering    
    \includegraphics[width = 0.95\linewidth]{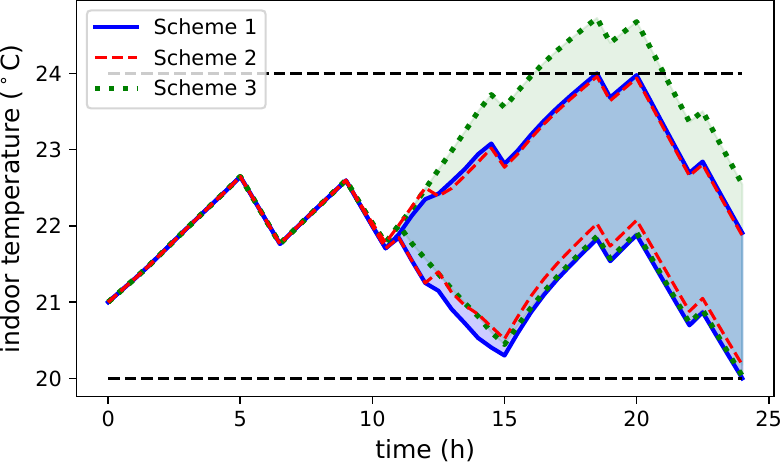}
    \caption{Indoor temperature envelopes for all schemes.}
    \label{fig:temp}
\end{figure}

Fig. \ref{fig:time_value} gives the computation time and the optimal $\Gamma^*$, i.e., the largest number of flexible switches of $\mathbf{r}_j$ $(j\in\mathcal{U})$, for all schemes with an increasing size of $|\mathcal{U}|$. Clearly, it can be observed from Fig. \ref{fig:time_value}a that the computation time of Scheme 2 (exhaustive search) increases exponentially with $|\mathcal{U}|$. On the contrary, our proposed approach is much more computationally efficient, and its computation time increases linearly with $|\mathcal{U}|$. Compared with Scheme 1, Scheme 3, which only considers open-loop control actions instead of closed-loop control policies as in Scheme 1, is computationally less demanding due to the lower number of decision variables. Obviously, the downside of Scheme 3 is its suboptimality, as shown in Fig. \ref{fig:time_value}b. In addition, we can also conclude from Fig. \ref{fig:time_value}b that, for this case study, the affine control policy \eqref{eq:control_policy} restricted in our proposed approach (Scheme 1) does not compromise the optimality of the solution and can find the optimal $\Gamma^*$ as with Scheme 2.

Fig. \ref{fig:temp} depicts the indoor temperature envelopes with all 3 schemes under all possible uncertain scenarios of $\bf r$ given $|\mathcal{U}| =8$ and $\Gamma = 5$. Clearly, Scheme 3 is unable to robustly guarantee indoor comfort constraints, which validates the conclusion that open-loop control actions are less robust than closed-loop control policies.


\section{CONCLUSIONS}\label{sec:5}
This paper investigated a class of ROC problems with binary adjustable uncertainties, that offers a framework for addressing the reserve and provision problem for energy systems with on-off type devices. In contrast to the existing methods for ROC with continuous adjustable uncertainties, a novel metric is introduced to measure the extent of binary uncertainties, and a general design framework for ROC with binary adjustable uncertainties is formulated. Subsequently, we reformulated and relaxed the ROC formulation by introducing auxiliary variables and applying strong duality, transforming it into a MILP problem amenable to efficient numerical solvers. This innovative approach empowers us to quantitively evaluate the flexibility potential inherent in modern energy systems with on-off type equipment. The efficacy of our proposed methodology is demonstrated through numerical experiments.

Future extensions include considering nonlinear constraints and nonaffine control policies, as well as hybrid types of adjustable uncertainties. 

\bibliographystyle{IEEEtran.bst}

\bibliography{ref}

\end{document}